\documentclass[prl,aps,floats,superscriptaddress,floatfix,twocolumn,bibliography]{revtex4-1}
 \usepackage{amssymb,amsmath}
\usepackage{amsmath,amssymb}
\usepackage{graphicx}
\usepackage{psfrag}

\usepackage{color}

\usepackage{float}
\restylefloat{table}

\usepackage{lipsum}
\usepackage[normalem]{ulem}
\usepackage{gensymb}

\newcommand{\beq}{\begin{equation}}
\newcommand{\eeq}{\end{equation}}
\newcommand{\beqn}{\begin{eqnarray}}
\newcommand{\eeqn}{\end{eqnarray}}

\def\beq{\begin{equation}}
\def\eeq{\end{equation}}
\def\bea{\begin{eqnarray}}
\def\eea{\end{eqnarray}}

\begin{document}
 \title{Flow can order: Phases of live XY spins in two dimensions}
\author{Astik Haldar}\email{astik.haldar@gmail.com}
\affiliation{Condensed Matter Physics Division, Saha Institute of
Nuclear Physics, 1/AF Bidhannagar, Calcutta 700064, West Bengal, India}
\author{Abhik Basu}\email{abhik.123@gmail.com}
\affiliation{Condensed Matter Physics Division, Saha Institute of
Nuclear Physics, 1/AF Bidhannagar, Calcutta 700064, West Bengal, India} 
\affiliation{Max-Planck Institut f\"ur Physik Komplexer Systeme, N\"othnitzer 
Str. 38,
01187 Dresden, Germany}

\date{\today}
\begin{abstract}
We present the hydrodynamic theory of active XY spins coupled with flow fields, for systems both having and or lacking number conservation in two dimensions (2D). For the latter, with strong activity or nonequilibrium drive, the system can synchronize, or be phase-ordered with various types of order, e.g., quasi long range order (QLRO) or new kind of order weaker or stronger than QLRO for sufficiently strong active flow-phase couplings. For the number conserving case, the system can show QLRO or order weaker than QLRO, again for sufficiently strong active flow-phase couplings. For other choices of the model parameters, the system necessarily disorders in a manner similar to immobile but active XY spins, or 2D Kardar-Parisi-Zhang surfaces.

\end{abstract}
\maketitle
 {\em Introduction:-}
  Systems out of equilibrium are often marked by their striking ability to display ordered states that are impossible in their equilibrium counterparts. For instance, a two-dimensional (2D) collection of self-propelled, orientable particles, known as a flock, can display long range orientational order in the presence of finite noises (nonequilibrium equivalent of temperature) and in the absence of any long-range interactions or symmetry-breaking external fields~\cite{toner-tu}. On the other hand, nonequilibrium effects can also completely disorder a system that otherwise shows order in the equilibrium limit~\cite{toner-cgle}.

In this Letter, we formulate the hydrodynamic theory of a collection of  mobile nearly phase-ordered oscillators, or ``active XY model'' with their velocity being redirected
by force densities arising from inhomogeneities in the phase in 2D, with or without number conservation. We show that this system is distinctly different from both its non-moving but active and equilibrium analogs. Our theory forms the basis of further studies in ordered states in woder ranging 2D systems with broken continuous symmetries, e.g., active superfluids~\cite{active-super1,active-super2,active-super3} (for which the entropy density is no longer a conserved hydrodynamic variable due to the activity) and oscillating chemical reactions~\cite{bz}. 

We restrict ourselves to a system with momentum conservation, but with or without number conservation. In the former case, the system is assumed to be incompressible, i.e., with a constant number density. In the latter case, we allow ``birth'' and ``death'' (hereafter the
“Malthusian” case~\cite{John-malth}), in the spirit of having ``live'' XY spins. In this case, the system is compressible, but the density fluctuations relax {\em fast} with a wavevector independent damping to a constant mean value determined by the birth and death rates. This means the local phase and the hydrodynamic velocity field are the only two hydrodynamic or slow variables in both the Malthusian and number conserving cases. 

We are interested in the variance $\Delta$ of the local phase fluctuations $\phi$ in the limit when the dynamics is dominated by the active processes: $\Delta \equiv\langle \phi^2({\bf x},t)\rangle$ in 2D. Our principal results are as follows.   For a range of the model parameters corresponding to sufficiently strong flow-phase coupling of generic active origin, (i) in the Malthusian case, we find a novel kind of ordered state that can be stable hydrodynamically and robust against noise:  the  variance $\Delta$ of the local phase fluctuations grow with the system size $L$ as $(\log L)^\mu$, where $\mu >0$ is a {\em nonuniversal exponent}, reflecting a slow growth with $L$. The exponent $\mu$ can be smaller or larger than unity; $\mu=1$ corresponds to quasi-long-range order (QLRO), also found in the 2D equilibrium XY model~\cite{chaikin}, $0<\mu<1$ ($\mu>1$) implies slower (faster) than logarithmic growth with $L$ that we name {\em stronger than QLRO} (SQLRO) ({\em weaker than QLRO} (WQLRO)). (ii) In the number-conserving incompressible case, the dominant active effects come from the chirality of the XY spins that couples the phase with the vorticity of the flow. For a range of the active parameters again $\Delta\sim (\log L)^\mu,\mu \geq 1$, indicating WQLRO. Remarkably, for other ranges of the active model parameters in both the Malthusian and the number conserving cases, the spins disorder as $L$ grows beyond a microscopic size, in a manner reminiscent of the roughening of 2D growing Kardar-Parisi-Zhang surfaces~\cite{kpz,stanley}. Thus appropriate flow-phase coupling can induce either order of various types in a collection of active XY spins, or disorder them.

We find that the nonuniversal exponent $\mu$ in the Malthusian case can in general depend on two dimensionless parameters $\alpha_1,\alpha_2$, both nonuniversal themselves. These are the ratios of two effective flow-phase coupling constants with another that represents the strength of the lowest order phase-phase coupling that exists even for immobile active XY spins. Out of these two, $\alpha_1$ is proportional to the ratio of the birth and death rates, can be of either sign and vanishes in the number conserving case; whereas $\alpha_2\geq 0 $ that models how phase fluctuations are influenced by the fluid vorticity, which is a number conserving process and can be present even in the non-Malthusian case. 

\begin{figure}[htb]
 \includegraphics[width=5cm]{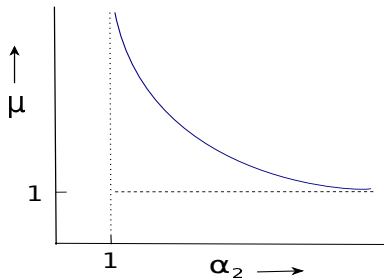}\hfill
\caption{Plot of $\mu$ as a function of $\alpha_2$ in the number conserving case.}\label{muplot1}
\end{figure}

In this Letter, for simplicity, we study the Malthusian case with $\alpha_2=0$ and obtain $\mu$ as a function of $\alpha_1$ that suffices our purpose here. The full form of $\mu$ as a function of both $\alpha_1$ and $\alpha_2$ are given in the associated long paper (ALP). { In the Malthussian case flow induced ordering takes place outside the window $1.3829> \alpha_1 >-0.1607$. WQLRO is found for $1.3829<\alpha_1 <2$ and $-1<\alpha_1 <-0.1607$, QLRO is for $\alpha_1=-1,2$. Else SQLRO is displayed for either $\alpha_1 >2$ or $\alpha_1 <-1$}. The nonlinear dependence of $\mu$ on $\alpha_1$ is shown in Fig.~\ref{muplot1} (a). The system disorders for $1.3829 > \alpha_1 >-0.1607$, that closely resembles the rough phase of  KPZ surface.
\begin{figure}[htb]
 \includegraphics[width=6cm]{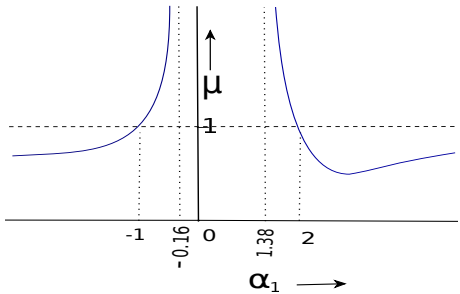}
 \caption{Plot of $\mu_1$ as a function of $\alpha_1$ for mathusian case}\label{muplot}
\end{figure}
For the number conserving incompressible case, WQLRO is obtained for $\alpha_2>1$, when $\mu=1+1/(\alpha_2-1)$ is necessarily larger than unity, as it should be for WQLRO; see Fig.~\ref{muplot1} for the nonlinear dependence of $\mu$ on $\alpha_2$. For $0<\alpha_2<1$,  disordered phase akin to the rough phase of a KPZ surface ensues.


We now outline the derivation of these results. Since we are considering an active or nonequilibrium system, we must begin by setting up the dynamical equations for the slow variables $\phi$ and $\bf v$. Symmetry considerations (invariance under translation and rotation in space, and an arbitrary but constant shift in the phase) require that the dynamical equation for $\phi$ to lowest order in a gradient expansion, take the form
\begin{eqnarray}
 \frac{\partial \phi}{\partial t} +\frac{\lambda}{2} ({\boldsymbol\nabla} \phi)^2 +\lambda_1 {\bf v}.{\boldsymbol\nabla} \phi &=& \nu \nabla^2 \phi + \lambda_2({\boldsymbol\nabla} \times {\bf v})_z \nonumber \\&+& \lambda_3 {\boldsymbol\nabla}\cdot {\bf v} + \xi ,\label{phieq}
\end{eqnarray}
Here, $\nu >0$ is the damping coefficient, $\lambda_1,\lambda_2,\lambda_3$ couple flow ($\bf v$) with $\phi$ and vanish for immobile oscillators (${\bf v}=0$), $\lambda$ is another coupling constant for a symmetry-permitted  nonlinearity that survives for immobile oscillators; $\xi$ is a Gaussian-distributed spatio-temporally white noise (since $\phi$ is non-conserved) with zero mean and a variance
\begin{equation}
 \langle \xi({\bf x},t) \xi(0,0)\rangle = 2D_0\delta^2({\bf x})\delta(t).\label{phi-noise}
\end{equation}
Terms with coefficients $\lambda,\lambda_2,\lambda_3$ are forbidden in equilibrium by the rotation invariance (invariance under a constant phase shift) of the underlying free energy in equilibrium. They are, however, permitted here  simply because rotation invariance
at the level of the equations of motion, which
is all that can be demanded in an active system, does not rule
them out; see the associated long paper (ALP) for a detailed derivation.
Coupling $\lambda,\,\lambda_2,\lambda_3$ can be of any sign;  Lastly, $\lambda_2,\lambda_3$ couple local change in $\phi$ with the local vorticity and compressibility of the fluid; for an achiral system $\lambda_2=0$ identically.

Assuming low Reynolds number flows, we ignore inertia of the fluid. Similar symmetry considerations lead to the generalized Stokes equation:
\begin{equation}
 \eta_1 \nabla^2 v_i +\eta_2 \nabla_i({\boldsymbol\nabla}\cdot {\bf v})= \nabla_i \varPi + w \nabla_i \phi \nabla^2\phi, \label{stokes}
\end{equation}
consistent with the momentum conservation; { we have ignored any external force in (\ref{stokes})}.
Here $\eta$ is the 2D fluid viscosity, $\varPi$ the pressure; $w$ is a coupling constant that controls how the local velocities are redirected by the local inhomogeneities in the phase; in the equilibrium limit, $w=\lambda_1$ as a condition for equilibrium. For an active system as here, $w$ and $\lambda_1$ are in principle free parameters. 
Equations (\ref{phieq}) and (\ref{stokes}) clearly are invariant under the Galilean transformation ${\bf v}\rightarrow {\bf v} + {\bf v}_0$. Finally, all terms in (\ref{phieq}) and (\ref{stokes}), except the term with coefficient $\lambda_2$ generalize straightforwardly to any arbitrary dimension; the term with coefficient $\lambda_2$ can exist only in 2D and models chiral effects. For achiral systems, $\lambda_2$ vanishes. Note also that with $\bf v=0$ as appropriate for immobile oscillators, (\ref{phieq}) reduce to the well-known KPZ equation~\cite{kpz}. Hence, (\ref{phieq}) and (\ref{stokes}) can also be interpreted as the coupled equations for a growing KPZ surface coupled to a flow field that in turn is affected by the local height fluctuations. However, there is a crucial difference: For a growing surface, $\phi$ is a single-valued height field and unbounded.  Clearly, states
with different heights are always physically distinguishable. In contrast, for phase oscillators $\phi$ is periodic and hence can support stable topological defects, whose unbinding in equilibrium is described by the well-known Kosterlitz-Thouless theory. Since we are considering nearly phase-ordered states, the number of such defects is implicitly assumed to vanish.

In the Malthusian case, the dynamical equation for density fluctuations $\delta\rho$, a nonhydrodynamic variable, with a
source term to reflect the tendency of birth and death
to restore the local population density to its steady state
value, and an additional, non-number-conserving Gaussian-distributed noise $f_\rho$
reflecting statistical fluctuations reads after dropping irrelevant terms~\cite{John-malth} and linearizing about  a mean density $\rho_0$
\begin{equation}
 \delta\rho =\gamma{\boldsymbol\nabla}\cdot {\bf v} + f_\rho,\label{deneq}
\end{equation} 
where $\gamma$ is a time-scale, a constant; see ALP for details. Further, applying an equation of state $\varPi(\delta\rho)=\psi\delta\rho$ locally, where $\psi$ is a susceptibility, (\ref{stokes}) reduce to
\begin{equation}
  \eta_1 \nabla^2 v_i +\eta_2 \nabla_i ({\boldsymbol\nabla}\cdot {\bf v})=  w \nabla_i \phi \nabla^2\phi + f_i,\label{stokes1}
\end{equation}
where $\psi$ and $f_\rho$ have been absorbed in $\eta_2$ and $f$, respectively. In the incompressible number conserving case, $\lambda_3=0$ in (\ref{phieq}) and ${\boldsymbol\nabla}\cdot {\bf v}=0$ in (\ref{stokes}).

We intend to calculate the universal scaling exponents that characterize the time-dependent correlation function of $\phi$, define as
\begin{equation}
 \langle \phi ({\bf x},t)\phi(0,0)\rangle \sim |x|^{2\chi}\theta(|x|^z/t),
\end{equation}
where $\chi$ and $z$ are, respectively, the roughness and dynamic exponent; $\theta$ is a dimensionless scaling function of its argument. In the linear passive theory ($\lambda=\lambda_1=\lambda_3=w=0$), $\chi$ and $z$ are known exactly: $\chi=0$ and $z=2$ in 2D. This corresponds to quasi-long range order with $\Delta\sim \log L$.

Simple power counting shows that the nonlinear terms are marginal at 2D. Thus the nonlinear effects can affect the scaling. To study this systematically, a perturbative dynamic renormalization group (RG) treatment is needed.

Couplings $\lambda,\,\lambda_1,\,\lambda_2,\,\lambda_3$ are all dimensionless; $\lambda_1$ can be set to 1.
All the active coefficients $\lambda,\,\lambda_2,\,\lambda_3$ must scale with the strength of the underlying nonequilibrium processes, or the energy released by them that ultimately drive the system away from thermal equilibrium. 
{In order to extract the role of the active effects on the phases in a systematic manner, we assume strong activity limit with $\lambda_1 \ll \lambda,\,\lambda_2,\,\lambda_3 $; further, we ignore any stochastic force $f_i$ in (\ref{stokes1}) for large $\eta_1,\, \eta_2$, consistent with the Stokesian limit for $\bf v$ .} 

To  proceed further, we note that the Stoksian velocity $\bf v$ that appears linearly in (\ref{stokes}) can be eliminated {\em exactly} to obtain an effective equation for $\phi$. The resulting equation, whose detailed form is given in ALP, now contains three nonlinear terms: one with coefficient $\lambda$ that is already existing in (\ref{phieq}), and two other equally relevant nonlinearities with coefficients $\tilde\lambda_2,\tilde\lambda_3$, that have its origin in the vorticity and compression terms with coefficients $\lambda_2$ and $\lambda_3$, respectively, in (\ref{phieq}).  As usual, the RG is done by tracing over
the short wavelength Fourier modes of the fields, followed by
a rescaling of lengths, times and the fields. This leads to the following
differential recursion relations:
\begin{eqnarray}
  &&\frac{d\nu}{dl} = \nu\left[z-2+(\frac{\alpha_1^2}{2}-\frac{5\alpha_1}{8}+\frac{\alpha_2}{8})g\right],\label{nu}\\
 &&\frac{dD}{dl} = D\left[z-2-2\chi+(\frac{1}{4}+\frac{3\alpha_1^2}{8}-\frac{\alpha_1}{2}+\frac{\alpha_2}{8})g\right],\label{D}\\
  &&\frac{d\lambda}{dl} = \lambda\left[z+\chi-2\right] ,\label{lambda}\\
  &&\frac{d \tilde\lambda_2}{dl} = \tilde\lambda_2 \left[z+\chi-2\right],\label{lambda_2}\\
  &&\frac{d \tilde\lambda_3}{dl}= \tilde\lambda_3\left[z+\chi-2\right],\label{lambda_3}
\end{eqnarray}
where we have defined two effective coupling constants $g=\lambda^2D/(2\pi\nu^3)$ and two dimensionless coupling constants $\alpha_1=\tilde\lambda_3/\lambda$ and $\alpha_2=\tilde\lambda_2^2/\lambda^2$. Here $\exp(l)$ is the length rescaling factor. Notice that by construction $g$ is non-negative, where as $\alpha_1,\,\alpha_2$ can be of any sign. At this stage, for reasons of simplicity, we extract the universal scaling for (i) $\alpha_2=0$ (i.e., $\lambda_2=0$) for Malthusian case and (ii) $\alpha_1=0$ (i.e., $\lambda_3=0$) for number conserving case separately.

(i) {\em Malthusian case}  ($\alpha_1\neq 0$): We set $\alpha_2=0$ in (\ref{nu}-\ref{D}) for simplicity.
The flow equations for $g,\,\alpha_1$ then read
\begin{eqnarray}
 &&\frac{dg}{dl} = \frac{g^2}{4}[1-\frac{9}{2}\alpha_1^2 + \frac{11\alpha_1}{2}]=-\frac{g^2}{8}{\cal A}_1 (\alpha_1),\label{flowmal}\\
 &&\frac{d\alpha_1}{dl}=0.\label{flowalpha1}
 \end{eqnarray}
 where ${\cal A}_1(\alpha_1)=9\alpha_1^2 - 11\alpha_1-2 >0$.
 Thus $\alpha_1$ is {\em marginal}. 
 For immobile oscillators, $\alpha_1=0$ identically, giving
 \begin{equation}
  \frac{dg}{dl} = \frac{g^2}{4}>0,
 \end{equation}
implying a run away flow, with $g$ diverging in ${\cal O}(1)$ renormalization group time $l$. This corresponds to a short range order or disorder, similar to the rough phase of a KPZ surface~\cite{kpz,natter,stanley}. Translating this result for phase-coupled oscillators, we find that nearly phase-ordered states of a collection of immobile oscillators are always unstable - they always {\em desynchronize}.

When the oscillators are mobile, $\alpha_1>0$, the physics can change dramatically. In fact, for appropriately chosen $\alpha_1$, if ${\cal A}_1(\alpha_1)>0$, then $g$ flows to zero in the long wavelength limit. The range of $\alpha_1$ for which this is possible is given by
\begin{equation}
 {\cal A}_1(\alpha_1)=9\alpha_1^2 - 11\alpha_1-2 >0.\label{a1}
\end{equation}
This holds so long as $\alpha_1 >(11+\sqrt{193})/18\approx 1.3829$, or $\alpha_1 < (11-\sqrt{193})/18\approx -0.1607$. Within the window $-0.1607<\alpha_1<1.3829$, ${\cal A}_1 <0$ and instability ensues. 

\begin{figure}[htb]
\includegraphics[width=5cm]{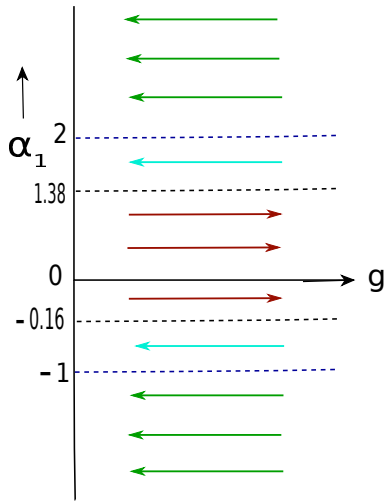}
\caption{Schematic flow diagram in the $g-\alpha_1$ plane. Stable and unstable regions are marked by the arrows indicating the direction of RG flow (see text).}\label{phase1}
\end{figure}

Various possibilities of ordered states emerge in the stable regime. In general in the ordered state  (\ref{flowmal}) gives
\begin{equation}
 g(l)=\frac{g(l=0)}{1+{\cal A}_1 l g(l=0)/8}\approx \frac{8}{{\cal A}_1 l},
\end{equation}
for $l\gg 1$. This gives, together with $z=2$ and $\chi=0$ and with the identification  $l=-\ln (a_0q)$, we get
\begin{equation}
 \nu(q)=C_\nu \left(\ln\frac{1}{q}\right)^{\Delta_1},\,D(q)=C_D\left(\ln\frac{1}{q}\right)^{\Delta_2},
\end{equation}
where $a_0$ is the lattice spacing, $\Delta_1=\frac{4\alpha_1^2-5\alpha_1}{{\cal A}_1(\alpha_1)},\Delta_2=\frac{3\alpha_1^2-4\alpha_1+2}{{\cal A}_1(\alpha_1)}$, and $C_\nu$ and $C_D$ are two dimensional constants. This gives
\begin{equation}
 \langle \phi^2 ({\bf x},t)\rangle = \int \frac{d^2 q}{(2\pi)^2}\frac{D(q)}{\nu(q)q^2}\sim (\ln L)^{1+\Delta_2-\Delta_1},
\end{equation}
{ giving $\mu=1+\Delta_2 -\Delta_1$. Therefore, QLRO ensues when $\Delta_1=\Delta_2$, i.e., $\alpha_1=2,\,-1$. Within the windows $1.3829<\alpha_1<2$ and $-1<\alpha_1 <-0.1607$, $\Delta_2-\Delta_1>0$, giving $\mu>1$. This means $\langle \phi^2 ({\bf x},t)\rangle$ diverges with $L$ faster than QLRO, demonstrating an order weaker than QLRO, that we call {\em weaker than QLRO} or {\em WQLRO}. On the other hand, for $\alpha_1>2$ or $\alpha_1<-1$, $\Delta_2-\Delta_1<0$, $\mu<1$ indicating $\langle \phi^2 ({\bf x},t)\rangle$ to grow with $\ln L$ slower than QLRO. This gives the name {\em stronger than QLRO} or {\em SQLRO}.Minimum of $\mu$ is 0.88805, obtained for $\alpha_1=15.347$ for which the divergence of $\langle \phi^2 ({\bf x},t)\rangle$ is slowest}.  For ${\cal A}_1<0$ [see Eq.~(\ref{a1}) above], the system disorders: We find
\begin{equation}
 \frac{dg}{dl}=|{\cal A}_1| g^2/8 \implies g(l)=\frac{g(l=0)}{1-|{\cal A}_1| g(l=0)/8}.
\end{equation}
 Thus $g$ diverges in a finite renormalization group time, or for a microscopic system size, implying disorder. This is reminiscent of roughening of 2D KPZ surfaces~\cite{stanley}. The RG
flow diagram in the $g-\alpha_1$ plane is given in Fig.~\ref{phase1}.

\begin{figure}[htb]
\includegraphics[width=5cm]{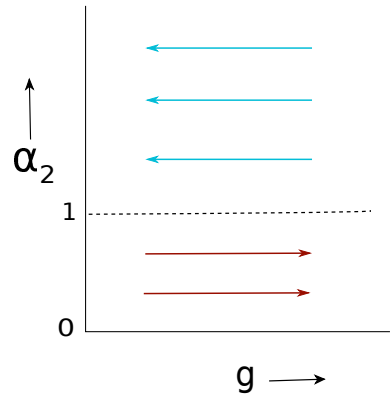}
 \caption{Schematic flow diagram in the $g-\alpha_2$ plane. Stable and unstable regions are marked by the arrows indicating the direction of RG flow (see text).}\label{rgflow2}
\end{figure}

(ii) {\em Number conserving case ($\alpha_1=0,\,\alpha_2 \neq 0$):-} This is relevant over time scales smaller $\gamma$ as defined in (\ref{deneq}). The differential recursion relations for $D$ and $\nu$ may be obtained from (\ref{nu}-\ref{D}) by setting $\alpha_1=0$. The corresponding
flow equations for $g, \alpha_2$ now read
\begin{eqnarray}
 &&\frac{dg}{dl}=\frac{g^2}{4}(1-\alpha_2)=-g^2{\cal A}_2 (\alpha_2),\label{flowcons}\\
 &&\frac{d\alpha_2}{dl}=0.\label{flowalpha2}
\end{eqnarray}
 Thus $\alpha_2$ is marginal; ${\cal A}_2=(\alpha_2-1)/4$.
Thus, for $\alpha_2>1$, $g(l)$ flows to zero for large $l$ corresponding to ordered states, whose scaling properties we calculate. We find
\begin{equation}
 g(l)=\frac{g(l=0)}{{\cal A}_2l g(l=0) + 1}\approx \frac{1}{{\cal A}_2l},
\end{equation}
for large $l$, where ${\cal A}_2>0$ for $\alpha_2>1$. On the other hand, when $\alpha_2<1$, ${\cal A}_2<0$ and hence there is a run away flow of $g$. This implies instability of the nearly phase ordered state, and is reminiscent of roughening of 2D KPZ surfaces. The RG flow lines are illustrated in Fig.~\ref{rgflow2}. Clearly $\alpha_2=1$ is a separatrix that separates the stable and unstable regions in the $g-\alpha_2$ plane; in fact, $\alpha_2=1$ is a {\em fixed line} on which $dg/dl=0$ identically.
As can be seen from that figure, flow lines with $\alpha_2=\tilde\lambda_2^2(l=0)/\lambda^2(l=0)>1$ flow towards the line $g=0$, where as flow lines with $\alpha_2<1$ flow away from the line $g(l)=0$.  Considering $\alpha_2>1$, i.e., above the separatrix in 
Fig.~\ref{rgflow2}, in the long wavelength limit $l\rightarrow \infty$, with $g (l)\sim \frac{1}{{\cal A}_2 l}$,


This gives together with $z=2,\,\chi=0$ with the identification $l\sim \ln(1/q)$, where $q$ is a Fourier wavevector, we obtain the scale-dependent renormalized damping $\nu(q)$ and noise variance $D(q)$
\begin{equation}
 \nu(q)=\nu_0[\ln(1/q)]^{\frac{\alpha_2}{8{\cal A}_2}},\;D(q)=D_0[\ln(1/q)]^{\frac{2+\alpha_2}{8{\cal A}_2}}.\label{nu-D-q}
\end{equation}
With (\ref{nu-D-q}) we now calculate $\Delta=\langle\phi^2({\bf x},t)\rangle$:
\begin{equation}
 \Delta \sim (\log L)^{1+ 1/(\alpha_2-1)}.\label{wqlro}
\end{equation}
For $\alpha_2>1$, $\Delta$ grows {\em faster} than just $\log L$ (expected for QLRO). Thus it is a weaker order than QLRO; we name it {\em weak QLRO} (WQLRO). Nonetheless, it is a {\em far stronger} order than just short range order (SRO) or disorder, where any notion of order is lost within a microscopic distance. As $\alpha_2\rightarrow\infty$, $\Delta\sim \log L$, recovering QLRO.

We now turn to the case $\alpha_2<1$. Then ${\cal A}_2(\alpha_2)<0$; hence
\begin{equation}
 \frac{dg}{dl}=|{\cal A}_2|g^2\implies g(l)=\frac{g(l=0)}{1-|{\cal A}_2|g(l=0)}.
\end{equation}
Thus, $g(l)$ {\em diverges} at a microscopic length scale $l_{crit}=1/[{\cal A}_2g(l=0)]$. Therefore, the system can remain ordered only if it is sufficiently small. This is again similar to the roughening of 2D KPZ surfaces.

The above results for a free standing system show that $\langle\theta^2 ({\bf x},t)\rangle$ steadily albeit slowly rises with $L$. Thus for a sufficiently large system size $L$, the nonlinear term $\lambda_1 {\bf v}\cdot {\boldsymbol\nabla} \theta$ should become relevant. The scale beyond which this happens should however be exponentially large in $\lambda_2$ or $\lambda_3$; see ALP.  On the other hand, the assumption of a 2D free standing system is of course admittedly an idealization; this breaks down when ``new physics'' not contained in the hydrodynamic model (\ref{phieq}) and (\ref{stokes}) intervenes beyond some length scale $L_\eta$. 

We now consider what ``new physics'' might intervene. In reality, a 2D system like the one considered here is resting on a three-dimensional (3D) bulk fluid or a bulk solid. As shown in the ALP, interactions with the surrounding bulk medium (fluid or solid) change the $\eta q^2 {\bf v}({\bf q},t)$ viscous term in (\ref{stokes}) to $\eta'|q|{\bf v}({\bf q},t)$ for a surrounding bulk fluid, or $(1/\xi){\bf v}({\bf q},t)$ as the dominant damping terms in (\ref{stokes}) for $L>L_\eta$. Here, $\eta'$ is a 3D viscosity and $\xi$ is a friction coefficient. Scale $L_\eta$ can be estimated by equating 
$\eta q^2 {\bf v}({\bf q},t)$ with $\eta'|q|{\bf v}({\bf q},t)$, or $(1/\xi){\bf v}({\bf q},t)$. This gives $L_\eta = \eta/\eta'$ for a 3D bulk fluid, and $L_\eta = \sqrt{\xi\eta}$ for a solid substrate in contact.

As we have seen above, the term with coefficient $\lambda$ pushes the system to disorder and desynchronization, for both the Malthusian and number conserving cases. In contrast, compressibility and the vorticity terms with coefficients with coefficients $\lambda_1$ and $\lambda_2$, respectively, for the Malthusian and the number conserving cases, can counteract the $\lambda$-term and help stabilizing order.  However, for wavevectors $q<2\pi/L_\eta$, or equivalently, for system size $L>L_\eta$,  the nonlinear active flow - phase coupling gets weakened and is {\em subleading} (in a RG sense) to the nonlinear term with coefficient $\lambda$. This makes it unable to compete with the distabilizing effect due to the $\lambda$-term. To possibilities now exist. If $L_\eta > \xi_{NL}$ the nonlinear length, which is the smallest length scale big enough
to allow enough renormalization group “time” $l$ for $\nu(l)$ and $D(l)$ to acquire the logarithmic corrections in (\ref{nu-D-q}), then for $L> L_\eta >\xi_{NL}$, for which the stabilizing active flow-phase coupling ceases to become relevant, $\nu(q)$ and $D(q)$ are already renormalized as given in (\ref{nu-D-q}). This means that for $L>L_\eta$ or for wavevectors $q<2\pi/L_\eta$, the hydrodynamics of $\phi$ is described by the KPZ equation, {\em but} with a damping $\nu(q)$ and a noise variance $D(q)$ that diverge as in (\ref{nu-D-q}).

The new theory has a pseudo-Galilean invariance that enforces non-renormalization of $\lambda$. The new effective coupling constant $\tilde g(l)=\lambda^2 D(l)/\nu^3(l)$. Due to the singular nature of $\nu(q)$ and $D(q)$, there are no fluctuation corrections to them in a perturbative expansion. Clearly, $\tilde g(l)$ is {\em not} marginal at 2D for both the Malthusian and number conserving cases: Under na\"{i}ve rescaling, thus,
\begin{eqnarray}
 \frac{d\tilde g}{dl}&\sim& -{\cal M}\frac{\tilde g}{l},\label{newflow}
\end{eqnarray}
where ${\cal M}= {\cal A}_1$, or ${\cal A}_2$, respectively, for the Malthusian and number conserving cases. Thus, $\tilde g$
 decays (grows) with $l$ for ${\cal M}>(<)0$. 
Thus for sufficiently small $\tilde g(l)$, (\ref{newflow}) tells us that $\tilde g(l)$ will asymptotically vanish, maintaining WQLRO. Noting that $\tilde g(l=0)= g(l=\log L_\eta)$, this may be achieved by tuning $L_\eta$. The latter tuning can be made possible by adding active materials at the interface between the 2D system and 3D bulk,  which can allow ``slip'' at the interface, potentially weakening the interaction. 

If $L_\eta <\xi_{NL}$, the system does not have enough renormalization group time $l$ for $\nu$ and $D$ to renormalize. In that case, for a system with $L>L_\eta$ for wavevectors $q<2\pi/L_\eta$, the hydrodynamics of $\phi$ is described by the standard KPZ equation~\cite{stanley}, that in 2D only has a rough phase. Thus, the system disorders as $L$ exceeds a microscopic size.

Finally, the length $\xi_{NL}=\exp(l^*)$ may be estimated by setting ${\cal A}_{1,2}l^*g(l=0)\sim 1$, giving $\xi_{NL}\sim \exp(1/{\cal A}_{1,2}g(l=0))$.

{\em summary:-} We have developed the hydrodynamic theory for live XY spins. This theory predicts that when the active effects dominate, the system can show strong or weak QLRO scaling for certain choices of the active coefficients. For other choices, the system completely disorders as its size exceeds a finite value in a manner reminiscent of roughening of 2D KPZ surfaces. We consider both the Malthusian and the number conserving cases. These lead to the phase diagrams (\ref{flowalpha1}) and (\ref{flowalpha2}) for the Malthusian and the number conserving cases, respectively. These results should help us understand recent experiments on driven diffusive superfluids or active bacterial superfluids.

{\em Acknowledgements:-} A.H. and A.B thank the
Alexander von Humboldt Stiftung (Germany) for partial
financial support under the Research Group Linkage
Programme scheme (2016).


\bibliography{citeXY}

\begin{thebibliography}{11}%
\makeatletter
\providecommand \@ifxundefined [1]{%
 \@ifx{#1\undefined}
}%
\providecommand \@ifnum [1]{%
 \ifnum #1\expandafter \@firstoftwo
 \else \expandafter \@secondoftwo
 \fi
}%
\providecommand \@ifx [1]{%
 \ifx #1\expandafter \@firstoftwo
 \else \expandafter \@secondoftwo
 \fi
}%
\providecommand \natexlab [1]{#1}%
\providecommand \enquote  [1]{``#1''}%
\providecommand \bibnamefont  [1]{#1}%
\providecommand \bibfnamefont [1]{#1}%
\providecommand \citenamefont [1]{#1}%
\providecommand \href@noop [0]{\@secondoftwo}%
\providecommand \href [0]{\begingroup \@sanitize@url \@href}%
\providecommand \@href[1]{\@@startlink{#1}\@@href}%
\providecommand \@@href[1]{\endgroup#1\@@endlink}%
\providecommand \@sanitize@url [0]{\catcode `\\12\catcode `\$12\catcode
  `\&12\catcode `\#12\catcode `\^12\catcode `\_12\catcode `\%12\relax}%
\providecommand \@@startlink[1]{}%
\providecommand \@@endlink[0]{}%
\providecommand \url  [0]{\begingroup\@sanitize@url \@url }%
\providecommand \@url [1]{\endgroup\@href {#1}{\urlprefix }}%
\providecommand \urlprefix  [0]{URL }%
\providecommand \Eprint [0]{\href }%
\providecommand \doibase [0]{http://dx.doi.org/}%
\providecommand \selectlanguage [0]{\@gobble}%
\providecommand \bibinfo  [0]{\@secondoftwo}%
\providecommand \bibfield  [0]{\@secondoftwo}%
\providecommand \translation [1]{[#1]}%
\providecommand \BibitemOpen [0]{}%
\providecommand \bibitemStop [0]{}%
\providecommand \bibitemNoStop [0]{.\EOS\space}%
\providecommand \EOS [0]{\spacefactor3000\relax}%
\providecommand \BibitemShut  [1]{\csname bibitem#1\endcsname}%
\let\auto@bib@innerbib\@empty
\bibitem [{\citenamefont {Toner}\ and\ \citenamefont {Tu}(1995)}]{toner-tu}%
  \BibitemOpen
  \bibfield  {author} {\bibinfo {author} {\bibfnamefont {J.}~\bibnamefont
  {Toner}}\ and\ \bibinfo {author} {\bibfnamefont {Y.}~\bibnamefont {Tu}},\
  }\href@noop {} {\bibfield  {journal} {\bibinfo  {journal} {Physical Review
  Letters}\ }\textbf {\bibinfo {volume} {75}} (\bibinfo {year}
  {1995})}\BibitemShut {NoStop}%
\bibitem [{\citenamefont {E.~Altman}\ and\ \citenamefont
  {Toner}(2015)}]{toner-cgle}%
  \BibitemOpen
  \bibfield  {author} {\bibinfo {author} {\bibfnamefont {L.~C. S.~D.}\
  \bibnamefont {E.~Altman}, \bibfnamefont {L.~M.~Sieberer}}\ and\ \bibinfo
  {author} {\bibfnamefont {J.}~\bibnamefont {Toner}},\ }\href@noop {}
  {\bibfield  {journal} {\bibinfo  {journal} {Physical Review X}\ }\textbf
  {\bibinfo {volume} {5}} (\bibinfo {year} {2015})}\BibitemShut {NoStop}%
\bibitem [{\citenamefont {H.~M.~L\'opez}(2015)}]{active-super1}%
  \BibitemOpen
  \bibfield  {author} {\bibinfo {author} {\bibfnamefont {C.~D. H. A. E.~C.}\
  \bibnamefont {H.~M.~L\'opez}, \bibfnamefont {J.~Gachelin}},\ }\href@noop {}
  {\bibfield  {journal} {\bibinfo  {journal} {Physical Review Letters}\
  }\textbf {\bibinfo {volume} {115}} (\bibinfo {year} {2015})}\BibitemShut
  {NoStop}%
\bibitem [{\citenamefont {R.~Labouvie}\ and\ \citenamefont
  {Ott}(2016)}]{active-super2}%
  \BibitemOpen
  \bibfield  {author} {\bibinfo {author} {\bibfnamefont {S.~H.}\ \bibnamefont
  {R.~Labouvie}, \bibfnamefont {B.~Santra}}\ and\ \bibinfo {author}
  {\bibfnamefont {H.}~\bibnamefont {Ott}},\ }\href@noop {} {\bibfield
  {journal} {\bibinfo  {journal} {Physical Review Letters}\ }\textbf {\bibinfo
  {volume} {116}} (\bibinfo {year} {2016})}\BibitemShut {NoStop}%
\bibitem [{\citenamefont {S.~Guo}\ and\ \citenamefont
  {Cheng}(2018)}]{active-super3}%
  \BibitemOpen
  \bibfield  {author} {\bibinfo {author} {\bibfnamefont {Y.~P. a. X.~X.}\
  \bibnamefont {S.~Guo}, \bibfnamefont {D.~Samanta}}\ and\ \bibinfo {author}
  {\bibfnamefont {X.}~\bibnamefont {Cheng}},\ }\href@noop {} {\bibfield
  {journal} {\bibinfo  {journal} {Proceedings of the National Academy of
  Sciences (USA)}\ }\textbf {\bibinfo {volume} {115}} (\bibinfo {year}
  {2018})}\BibitemShut {NoStop}%
\bibitem [{\citenamefont {M.~Toiya}\ and\ \citenamefont {Epstein}(2010)}]{bz}%
  \BibitemOpen
  \bibfield  {author} {\bibinfo {author} {\bibfnamefont {V.~K. V. S.~F.}\
  \bibnamefont {M.~Toiya}, \bibfnamefont {H.~O. Gonz\'alez-Ochoa}}\ and\
  \bibinfo {author} {\bibfnamefont {I.~R.}\ \bibnamefont {Epstein}},\
  }\href@noop {} {\bibfield  {journal} {\bibinfo  {journal} {Journal of
  Physical Chemistry Letters}\ }\textbf {\bibinfo {volume} {1}} (\bibinfo
  {year} {2010})}\BibitemShut {NoStop}%
\bibitem [{\citenamefont {Toner}(2012)}]{John-malth}%
  \BibitemOpen
  \bibfield  {author} {\bibinfo {author} {\bibfnamefont {J.}~\bibnamefont
  {Toner}},\ }\href {\doibase 10.1103/PhysRevLett.108.088102} {\bibfield
  {journal} {\bibinfo  {journal} {Physical Review Letters}\ }\textbf {\bibinfo
  {volume} {108}},\ \bibinfo {pages} {088102} (\bibinfo {year}
  {2012})}\BibitemShut {NoStop}%
\bibitem [{\citenamefont {Chaikin}\ and\ \citenamefont
  {Lubensky}(2000)}]{chaikin}%
  \BibitemOpen
  \bibfield  {author} {\bibinfo {author} {\bibfnamefont {P.~M.}\ \bibnamefont
  {Chaikin}}\ and\ \bibinfo {author} {\bibfnamefont {T.~C.}\ \bibnamefont
  {Lubensky}},\ }\href@noop {} {\emph {\bibinfo {title} {Principles of
  condensed matter physics}}},\ Vol.~\bibinfo {volume} {1}\ (\bibinfo
  {publisher} {Cambridge university press Cambridge},\ \bibinfo {year}
  {2000})\BibitemShut {NoStop}%
\bibitem [{\citenamefont {Kardar}\ \emph {et~al.}(1986)\citenamefont {Kardar},
  \citenamefont {Parisi},\ and\ \citenamefont {Zhang}}]{kpz}%
  \BibitemOpen
  \bibfield  {author} {\bibinfo {author} {\bibfnamefont {M.}~\bibnamefont
  {Kardar}}, \bibinfo {author} {\bibfnamefont {G.}~\bibnamefont {Parisi}}, \
  and\ \bibinfo {author} {\bibfnamefont {Y.-C.}\ \bibnamefont {Zhang}},\
  }\href@noop {} {\bibfield  {journal} {\bibinfo  {journal} {Physical Review
  Letters}\ }\textbf {\bibinfo {volume} {56}},\ \bibinfo {pages} {889}
  (\bibinfo {year} {1986})}\BibitemShut {NoStop}%
\bibitem [{\citenamefont {Barab{\'a}si}\ and\ \citenamefont
  {Stanley}(1995)}]{stanley}%
  \BibitemOpen
  \bibfield  {author} {\bibinfo {author} {\bibfnamefont {A.-L.}\ \bibnamefont
  {Barab{\'a}si}}\ and\ \bibinfo {author} {\bibfnamefont {H.~E.}\ \bibnamefont
  {Stanley}},\ }\href@noop {} {\emph {\bibinfo {title} {Fractal concepts in
  surface growth}}}\ (\bibinfo  {publisher} {Cambridge university press},\
  \bibinfo {year} {1995})\BibitemShut {NoStop}%
\bibitem [{\citenamefont {L-H~Tang}\ and\ \citenamefont
  {Forest}(1990)}]{natter}%
  \BibitemOpen
  \bibfield  {author} {\bibinfo {author} {\bibfnamefont {T.~N.}\ \bibnamefont
  {L-H~Tang}}\ and\ \bibinfo {author} {\bibfnamefont {B.~M.}\ \bibnamefont
  {Forest}},\ }\href@noop {} {\bibfield  {journal} {\bibinfo  {journal}
  {Physical Review Letters}\ }\textbf {\bibinfo {volume} {65}} (\bibinfo {year}
  {1990})}\BibitemShut {NoStop}%
\end{thebibliography}%

\end{document}